\documentclass{PoS}

\newcommand{\ZZ}{Z \!\!\! Z}

\def\lsi{\raise0.3ex\hbox{$<$\kern-0.75em\raise-1.1ex\hbox{$\sim$}}}
\def\gsi{\raise0.3ex\hbox{$>$\kern-0.75em\raise-1.1ex\hbox{$\sim$}}}
\def\backder{\raise1.4ex\hbox{$\leftarrow$\kern-0.75em\raise-1.4ex\hbox{$\partial$}}}
\def\forder{\raise1.4ex\hbox{$\rightarrow$\kern-0.8em\raise-1.4ex\hbox{$\partial$}}}

\newcommand{\backderi}{\mathop{\backder}}
\newcommand{\forderi}{\mathop{\forder}}

\newcommand{\C}{{\kern+.25em\sf{C}\kern-.45em\sf{{\small{I}}} \kern+.45em\kern-.25em}}
\newcommand{\R}{{\kern+.25em\sf{R}\kern-.78em\sf{I} \kern+.78em\kern-.25em}}

\newcommand{\beq}{\begin{equation}}
\newcommand{\eeq}{\end{equation}}

\title{Simulation Results for $U(1)$ Gauge Theory on Non-Commutative Spaces}

\ShortTitle{Non-commutative U(1) gauge theory}

\author{\speaker{Wolfgang Bietenholz}%
\\
NIC/DESY Zeuthen, Platanenallee 6, D-15738 Zeuthen, Germany\\
        E-mail: \email{bietenho@ifh.de}}
\author{Antonio Bigarini\\
Dipartimento di Fisica, Universit\`{a} degli Studi di Perugia,
Via Pascoli 1, I-06100 Perugia, Italy\\
        E-mail: \email{Antonio.Bigarini@pg.infn.it}}
\author{Jun Nishimura\\
High Energy Accelerator Research Organization (KEK) and
Dept.\ of Particle and Nuclear Physics, 
Graduate University for Advanced Studies 
Tsukuba, Ibaraki, 305-0801, Japan\\
        E-mail: \email{jnishi@post.kek.jp}}
\author{Yoshiaki Susaki\\
High Energy Accelerator Research Organization (KEK) and
Graduate School of Pure and Applied Science, University of Tsukuba,
Tsukuba, Ibaraki, 305-0801, Japan\\
        E-mail: \email{susaki@het.ph.tsukuba.ac.jp}}
\author{Alessandro Torrielli\\
Center for Theoretical Physics, Laboratory for Nuclear Sciences 
and Dept.\ of Physics, Massachusetts Institute of Technology (MIT),
77 Massachusetts Avenue, MA 02139-4307, USA \\
        E-mail: \email{torriell@MIT.EDU}}
\author{Jan Volkholz\\
Institut f\"{u}r Physik, Humboldt-Universit\"{a}t zu Berlin,
Newtonstr. 15, D-12489 Berlin, Germany\\
E-mail: \email{volkholz@physik.hu-berlin.de}}

\abstract{We present numerical results for $U(1)$ gauge theory in 2d 
and 4d spaces involving a non-commutative plane. Simulations are feasible 
thanks to a mapping of the non-commutative plane onto a twisted 
matrix model. In $d=2$ it was a long-standing issue if Wilson loops are 
(partially) invariant under area-preserving diffeomorphisms. We show that 
non-perturbatively this invariance breaks, including the subgroup
$SL(2,R)$.
In both cases, $d=2$ and $d=4$, we extrapolate our 
results to the continuum and infinite volume by means of a Double 
Scaling Limit. In $d=4$ this limit leads to a phase with broken
translation symmetry, which is not 
affected by the perturbatively known IR instability. Therefore the 
photon may survive in a non-commutative world.}

\FullConference{The XXV International Symposium on Lattice Field Theory\\
                 July 30 - August 4 2007\\
                 Regensburg, Germany}

\begin{document}

\section{Non-commutative $U(1)$ gauge theory}

The positions in a non-commutative (NC) Euclidean plane correspond to
the spectra of Hermitian operators $\hat x_{1}, \  \hat x_{2}$,
with a non-vanishing commutator
\begin{equation}  \label{NCrel}
[\hat x_{\mu} ,\hat x_{\nu}] = i \Theta_{\mu \nu} =
i \theta \epsilon_{\mu \nu} \ .
\end{equation}
We treat the non-commutativity parameter $\theta$
as a constant. Relation (\ref{NCrel}) implies
a spatial uncertainty of the form
$\Delta x_{1} \Delta x_{2} \sim \theta$, which can be interpreted
as the event horizon of a strong gravitation centre. In fact, if it 
has been argued that attempts to merge quantum theory with
gravity lead quite generally to such a spatial uncertainty \cite{DFR},
which corresponds to a NC geometry.

In quantum field theory the spatial uncertainty gives rise to
non-locality over a range of $O(\sqrt{\theta})$. A related
consequence is the notorious ``UV/IR mixing'' \cite{MRS}: nested singularities
can be UV divergent in one momentum component $p_{\mu}$ and
IR divergent in another component $p_{\nu}$. Due to this property
the perturbative treatment is extremely involved.

Hence it is strongly motivated to take a fully non-perturbative approach.
As in commutative field theory it relies on the lattice regularisation
(for a review, see Ref.\ \cite{Szabo}).
Although we do not have sharp points as lattice sites, a (fuzzy)
lattice structure can be imposed by the operator identity
\begin{equation}  \label{expid}
\exp \Big( i \frac{2 \pi}{a} \hat x_{\mu} \Big) = \hat {1 \!\!\! 1} \ ,
\end{equation}
where $a$ is the lattice spacing.
Along with the usual periodicity of the momenta over the Brillouin
zone, identity (\ref{expid}) entails
$\frac{1}{2a} \theta p_{\mu} \in \ZZ \ .$
For fixed parameters $a$ and $\theta$ we infer that the lattice
is automatically {\em periodic,} in striking contrast to the commutative
lattice.

On a periodic $N \times N$ lattice one readily identifies
\begin{equation}
\theta = \frac{1}{\pi} N a^{2} \ .
\end{equation}
Therefore we extrapolate to the {\em Double Scaling Limit} (DSL)
\begin{equation}
\{ \quad a \to 0 \quad {\rm and} \quad N \to \infty \quad \}
\qquad {\rm at} \quad Na^{2} = {\rm const.}
\end{equation}
The DSL leads to a {\em continuous NC plane of infinite extent.}
The requirement to take the UV and IR limits simultaneously
in a controlled manner is again related to the UV/IR mixing.

We can return to 
ordinary coordinates $x_{\mu}$
if we multiply all fields by {\em star products,}
\beq
\phi (x) \star \psi (x) := \phi (x) \exp \Big( \frac{i}{2} 
\backderi \ \!\!\! _{\mu} \Theta_{\mu \nu} \! \forderi \ \!\!\! _{\nu} 
\Big) \psi (x) \ ,
\eeq
which encode the non-locality.
The star commutator 
$[ x_{\mu},x_{\nu}]_{\star} := x_{\mu} \star x_{\nu} -
x_{\nu} \star x_{\mu} =i \Theta_{\mu \nu}$
suggests that this transition is sensible --- it can 
be justified e.g.\ with a plane wave decomposition. 

In this framework we formulate $U(1)$ gauge theory on a NC plane as
\beq
S[A] = \frac{1}{4} \int d^{2} x \ F_{\mu \nu} \star F_{\mu \nu} \ , 
\quad
F_{\mu \nu} = \partial_{\mu} A_{\nu} - \partial_{\nu} A_{\mu} 
+ i g [ A_{\mu} , A_{\nu} ]_{\star} \ .
\eeq
Note that even the $U(1)$ gauge field picks up a self-interaction term. 
This action is invariant under star gauge
transformations. However, even on the lattice its direct
simulation is hardly feasible (for instance the 
compact formulation would require star unitary link variables).

We arrive at a numerically tractable formulation based on
the equivalence of this model with the twisted Eguchi-Kawai (TEK)
model \cite{TEK}. 
This matrix model is defined on one point with the action
\beq
S_{\rm TEK} [U] = -N \beta \sum_{\mu \neq \nu} Z_{\mu \nu} {\rm Tr}
[ U_{\mu} U_{\nu} U_{\mu}^{\dagger} U_{\nu}^{\dagger} ] \quad , \qquad
\beta \equiv 1 / g^{2} \ .
\eeq
The $U_{\mu}$ are unitary $N\times N $ matrices, which contain the
degrees of freedom of the lattice gauge field. The twist factor
$Z_{21} = Z_{12}^{*} = \exp( 2 \pi i n /N)$ makes the difference
from the original Eguchi-Kawai model --- in $d=4$ it avoids the 
spontaneous breaking of the centre symmetry at weak coupling.
$n$ is an integer, which 
we set $n = 
(N+1)/2$, so we always deal
with odd matrix/lattice sizes $N$.

Then the TEK can be identified with the NC $U(1)$ lattice gauge model
since the algebras are identical 
--- this has been shown in the large $N$ limit \cite{AIIKKT} 
and also at finite $N$ \cite{AMNS}.
Clearly, the TEK is numerically tractable, so that the simulations
can start if we also formulate suitable observables. In the matrix
model framework it is obvious to write down the analogue
of a Wilson loop,
\beq  \label{Wloop}
W_{\mu \nu} (I \times J) := \frac{1}{N} \, Z_{\mu \nu}^{ I \cdot J} \,
{\rm Tr} [ U_{\mu}^{I} U_{\nu}^{J} U_{\mu}^{\dagger \, I} 
U_{\nu}^{\dagger \, J} ] \ .
\eeq
Mapping this quantity back to the lattice yields indeed the
star gauge invariant term, which is considered the NC Wilson
loop \cite{NCWil}. This Wilson loop is complex in general,
$W_{\mu \nu} \in \C$, but the action is real since both orientations
are summed over (and $W_{\mu \nu} = W_{\nu \mu}^{*}$). Hence 
simulations are possible without running into a sign problem.
Such a study, including extrapolations to the DSL, was first
presented in Ref.\ \cite{BHN}. The planar limit coincides
with $U(N \to \infty )$ lattice field theory,
where the (real) Wilson loop follows an exact area law \cite{GroWi}.
Although this is not the limit that we are interested in,
it can be used to set the scale as 
$a^{2} = - \ln [1 - 1 / (4 \beta)]$ \ (for $\beta \geq 1/2$),
so that the DSL has to be taken roughly at a fixed 
ratio $N / \beta$.

\section{Wilson loops in $d=2$ : area-preserving diffeomorphisms (APDs)}

On the commutative plane, pure $U(n)$ gauge theories are
analytically soluble with geometric methods. 
Due to APD invariance, the expectation
values of Wilson loops only depend on the oriented area \cite{Wit}.

\begin{figure}[ht!]
\vspace*{-1mm}
\begin{center}
\vspace*{-2mm}
 \includegraphics[width=.2\linewidth,angle=0]{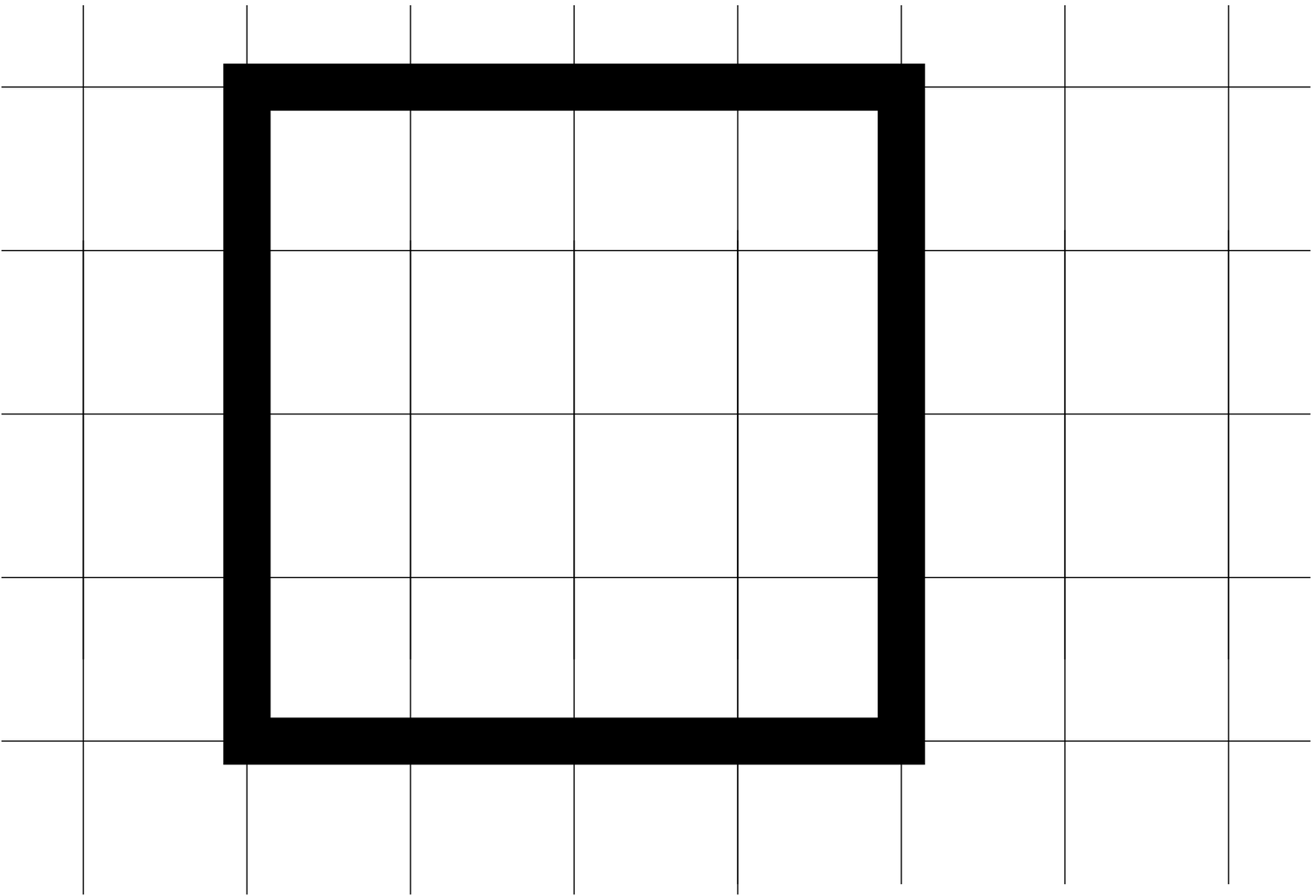} 
\hspace*{3mm}
 \includegraphics[width=.2\linewidth,angle=0]{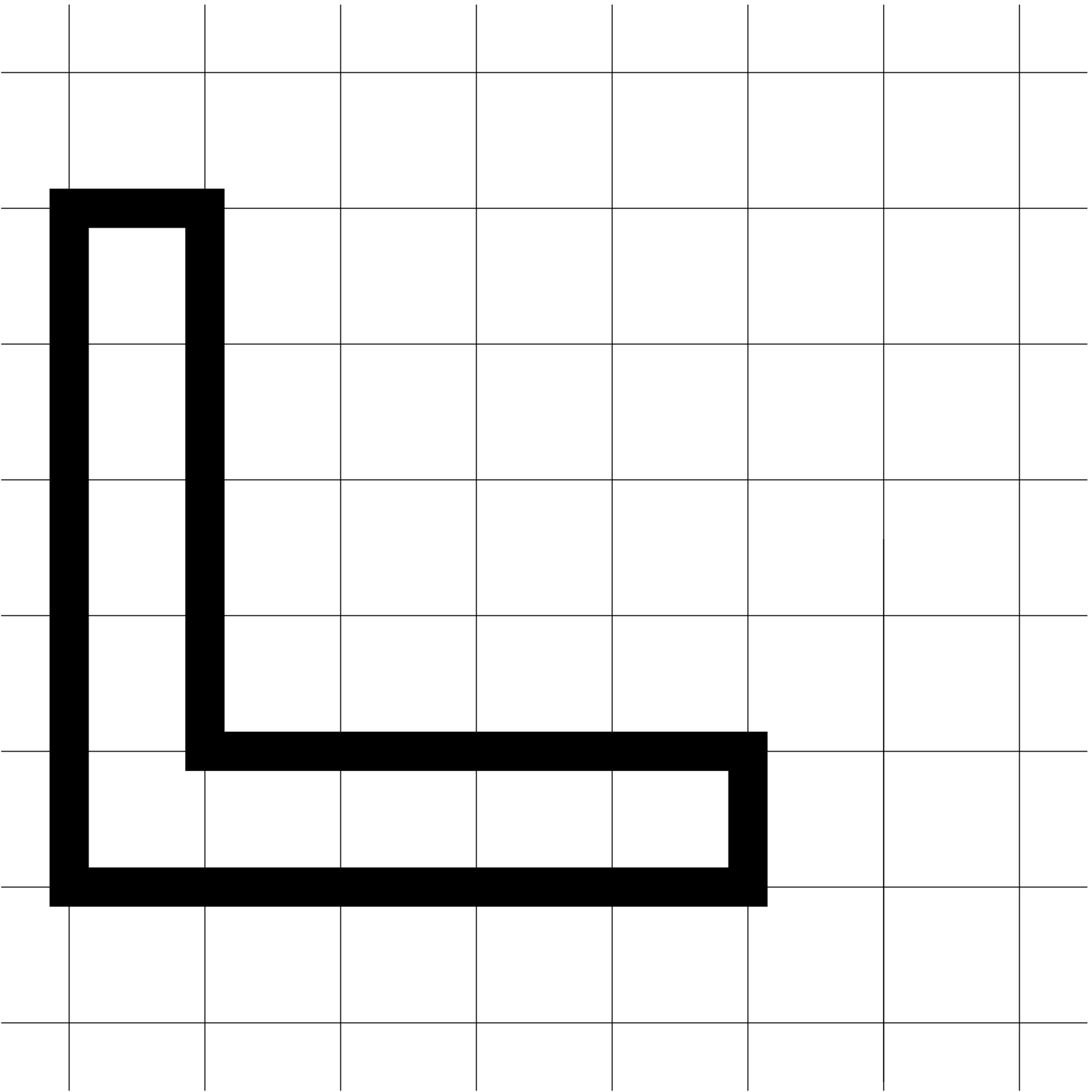}
\vspace*{3mm}
 \includegraphics[width=.2\linewidth,angle=0]{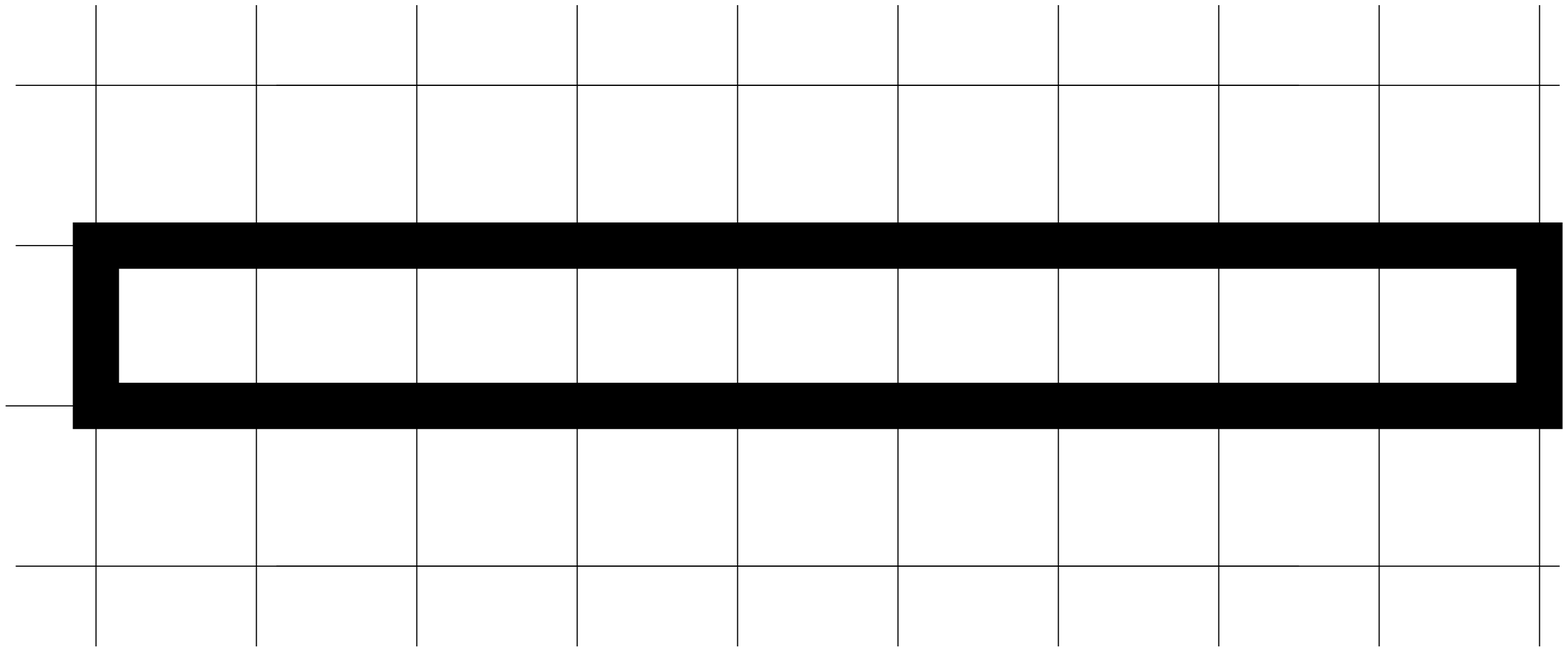} 
\hspace*{3mm}
 \includegraphics[width=.2\linewidth,angle=0]{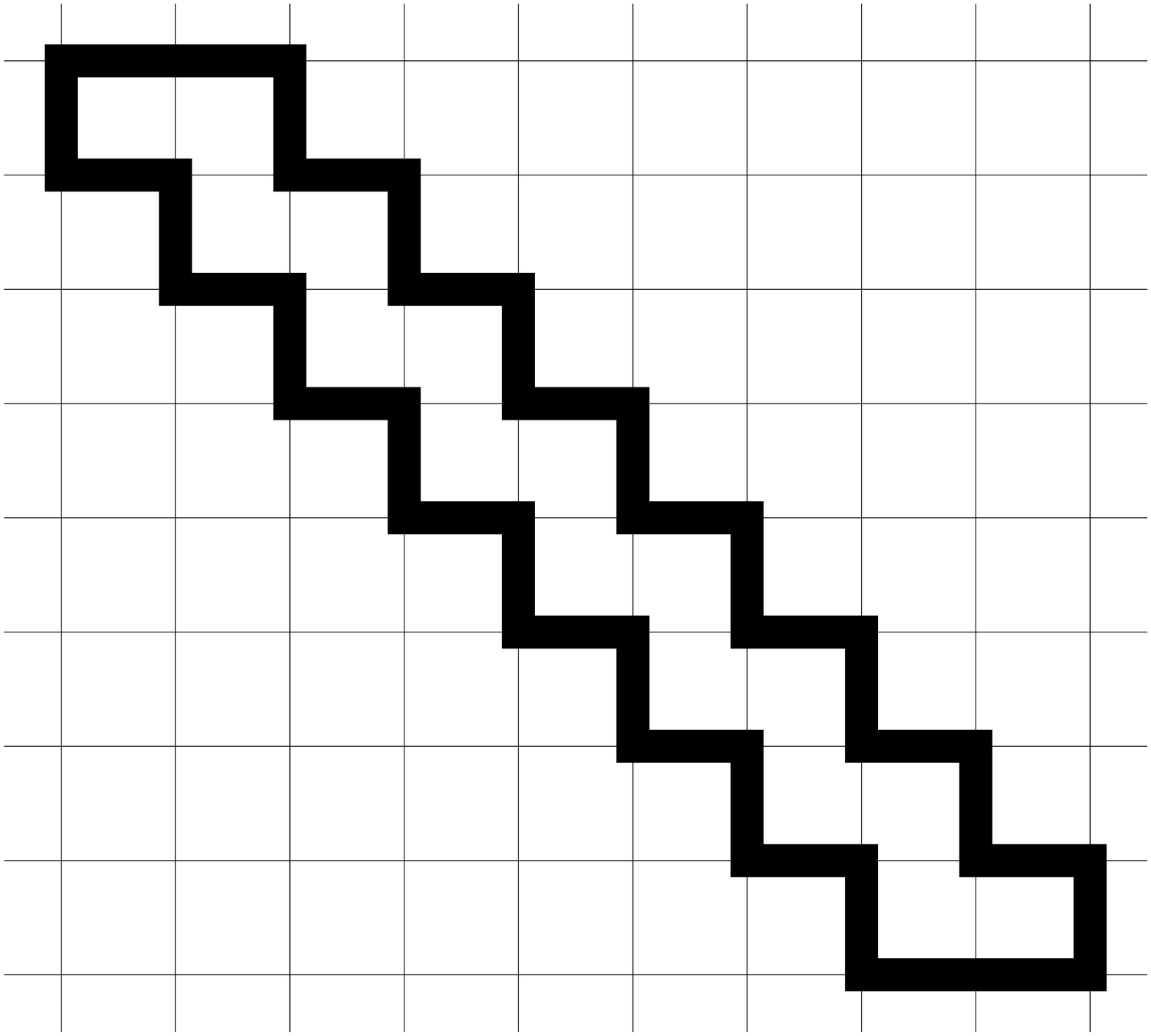}
\vspace*{-3mm} \\
\end{center}
\vspace*{-3mm}
\caption{\it{Examples for {\em squares, L-shapes, rectangles} and 
{\em stairs,} the four types of Wilson loops that we measured at the same
area to investigate the behaviour under APD transformation,
cf.\ Figure 2.}}
\vspace*{-3mm}
\label{shapes}
\end{figure}

\begin{figure}[ht!]
\begin{center}
\includegraphics[width=.35\linewidth,angle=270]{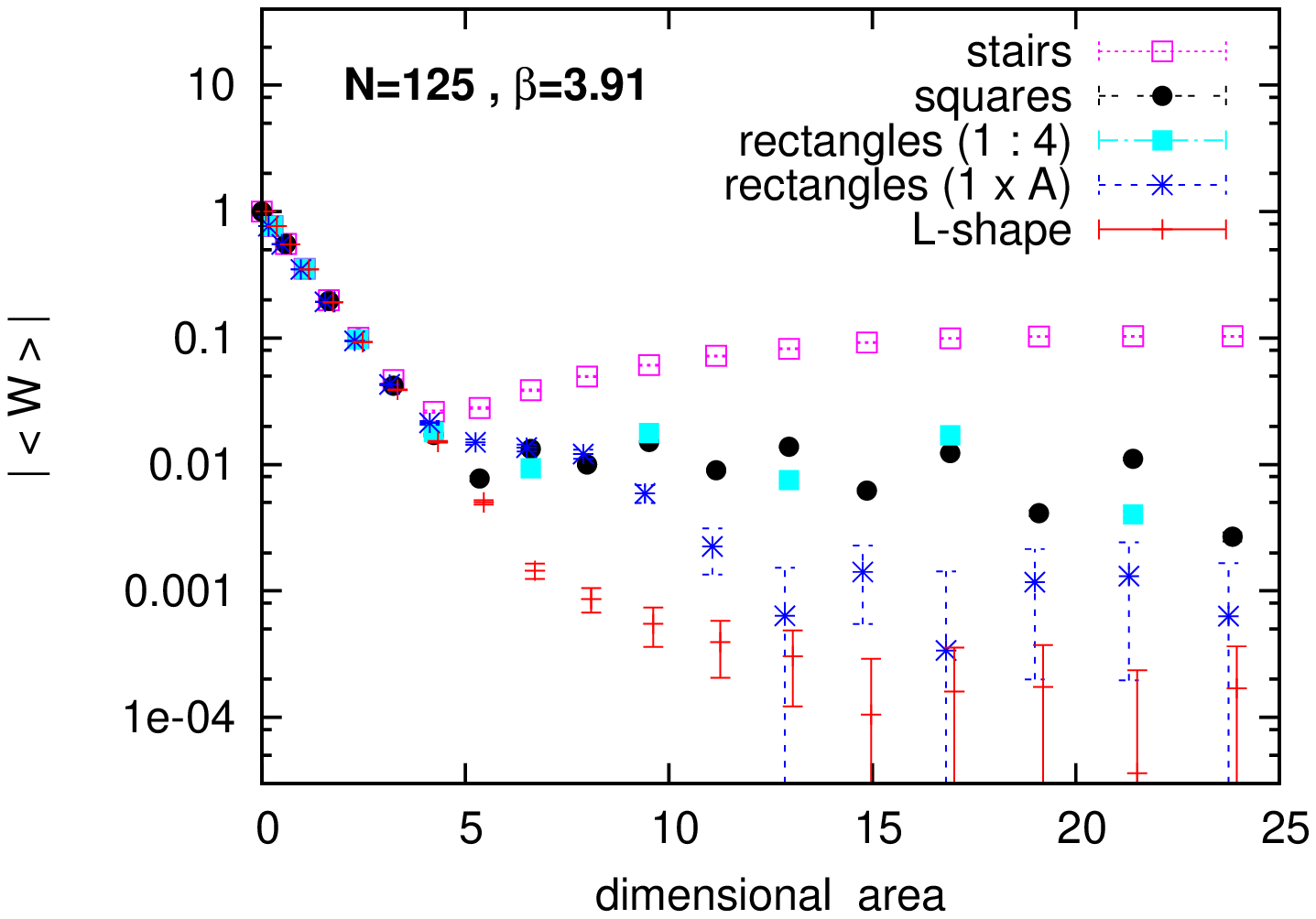}
\hspace*{-2mm}
\includegraphics[width=.35\linewidth,angle=270]{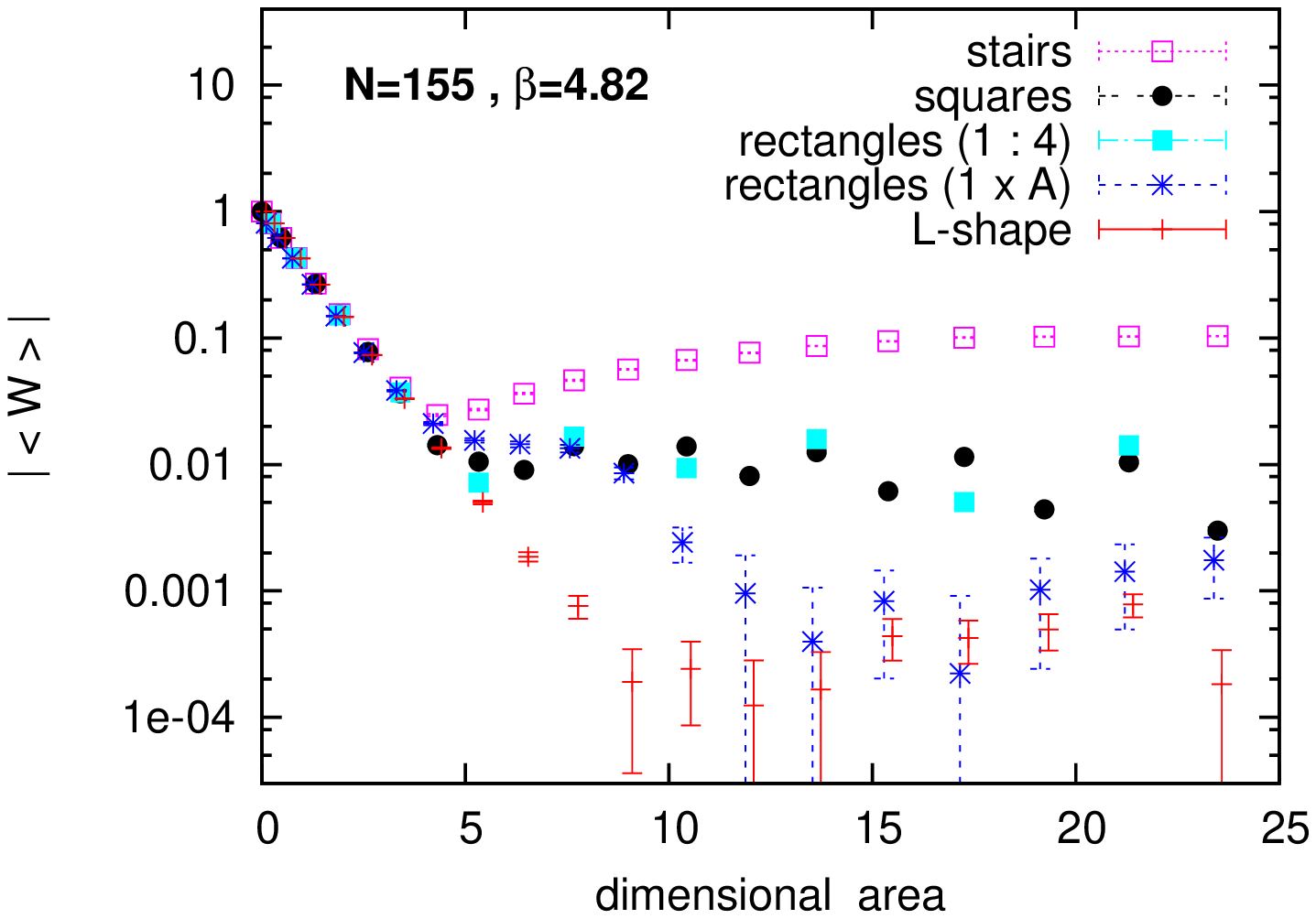}
\vspace*{-3mm} \\
\end{center}
\caption{{\it Absolute Wilson loop values of different shapes
for two volumes and lattice spacing at $\theta = 2.63$. The good
agreement between the two plots indicates that the APD
symmetry breaking persists in the DSL.}}
\label{Wloops1}
\vspace*{-2mm}
\end{figure}

Contrary to original expectations, it turned out that this symmetry
does not hold on the NC plane. In particular, perturbation theory
to $O(g^{4})$ and $O(\theta ^{-2})$ revealed that it breaks down
to $SL(2,R)$ \cite{pertu}, in agreement 
with other considerations \cite{semiclass}.
By means of numerical simulations we investigated the non-perturbative
extent of this APD symmetry breaking, as well as the viability of the
subgroup $SL(2,R)$ \cite{BBT}. To this end, we considered four types of
non-intersecting Wilson loops
with polygonal boundaries, generalising the form (\ref{Wloop}).
Prototypes are depicted in Figure \ref{shapes}.
Figure \ref{Wloops1} shows results for a set of Wilson loops at $N=125$, 
$\beta = 3.91$ (on the left) and $N=155$, $\beta = 4.82$ 
(on the right). In both cases the non-commutativity parameter amounts to
$\theta = 2.63$, hence we see two snapshots on the way to the
DSL: the right-hand-side corresponds to a larger volume with a finer
lattice. For a fixed (dimensional) loop area 
the results are very similar, hence we have apparently reached
the asymptotic DSL behaviour. At small areas we observe agreement
with the Gross-Witten area law \cite{GroWi}
(and therefore with the planar limit)
for all shapes. On the other hand, at
larger areas $| \langle W \rangle |$ does not decay further
and the values for the different shapes drift apart. This shows the
extent of APD symmetry breaking, which seems to persist in the
DSL to the continuum and infinite volume.
The rectangles (including squares) are related by
the APD subgroup 
$SL(2,R)$. In fact, $SL(2,R)$ has a higher viability
as an approximate symmetry subgroup --- at least up to moderate
deformations --- but it also breaks on the non-perturbative level.


\section{The fate of the photon in a non-commutative world}

\begin{figure}
\vspace*{-2mm}
\begin{center}
\includegraphics[angle=270,width=.34\linewidth]{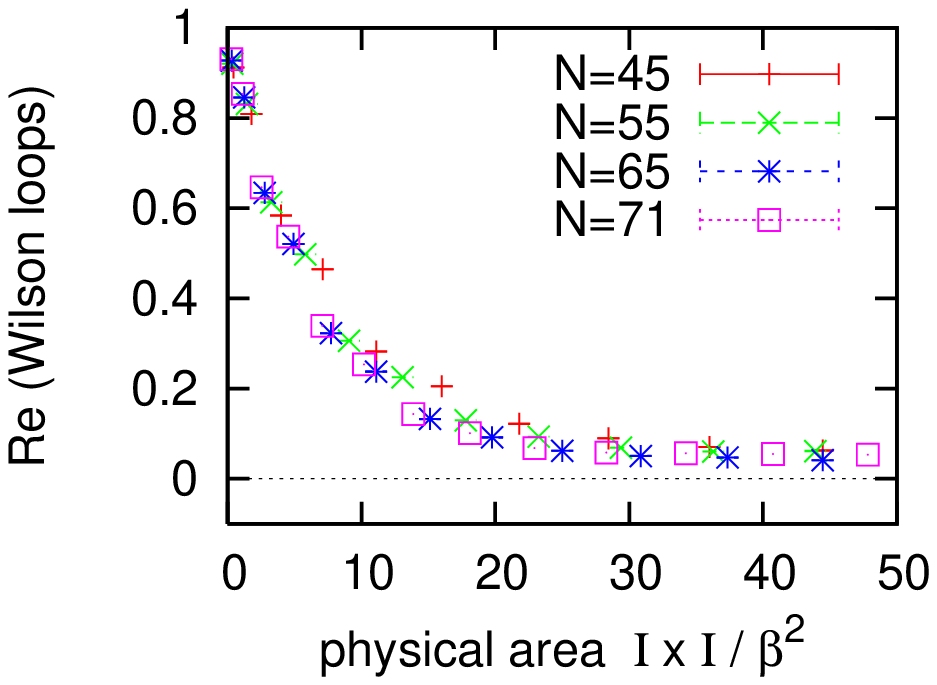} \hspace*{-5mm}
\includegraphics[angle=270,width=.34\linewidth]{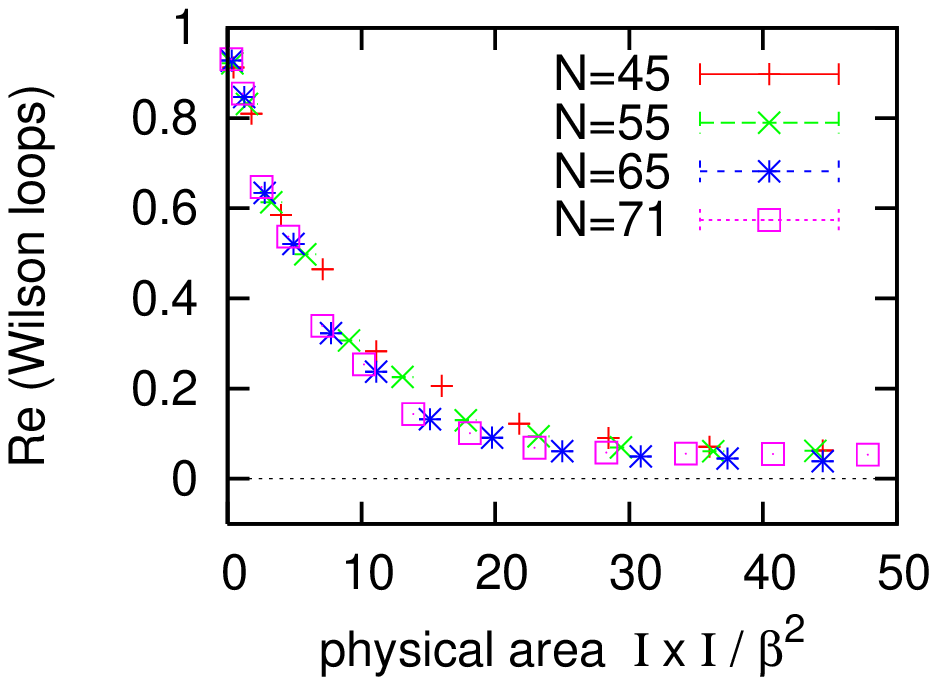} \hspace*{-5mm}
\includegraphics[angle=270,width=.34\linewidth]{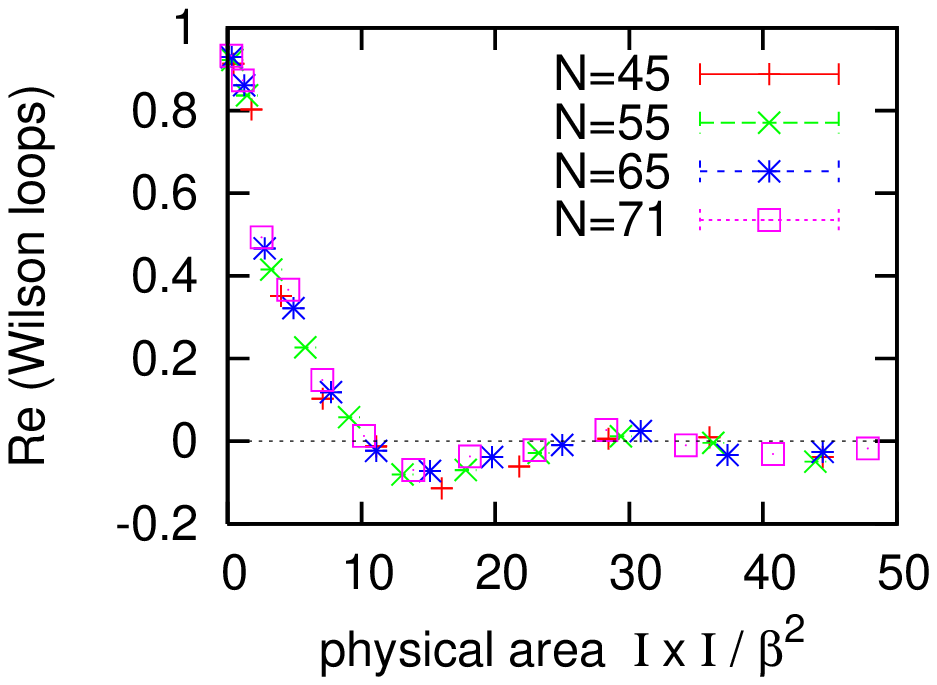}
\vspace*{-3mm} \\
\end{center}
\caption{{\it Results for Wilson loop in NC QED$_{4}$ at 
$N /\beta^{2} \equiv 20$ in the commutative plane, the mixed plane and
the NC plane (from left to right). In the first two planes
$\langle W \rangle$ is real, whereas the oscillation of 
${\rm Re} \, \langle W \rangle$ comes along with a rotating 
complex phase (similar to NC QED$_{2}$ \cite{BHN}).}} 
\label{4dWl}
\end{figure}

We consider again a NC plane with $[ \hat x_{1}, \hat x_{2}] = i \theta 
= {\rm const.}$, but now we add a commutative plane $(x_{3},x_{4})$,
which includes the Euclidean time. We are particularly interested
in a possible {\em $\theta$-distorted dispersion relation} of the photon,
which could in principle be experimentally measurable. 
A one loop calculation suggests 
the form \cite{MST}
\begin{equation}  \label{disp}
E^{2} = \vec p^{\, 2} + \frac{C}{\tilde p_{\mu} \tilde p_{\mu}} \qquad  
{\rm where}  \quad  C = {\rm const.} \ , \quad  \tilde p_{\mu} := 
\Theta_{\mu \nu} p_{\nu} \ .
\end{equation}
Corresponding phenomenological data have been analysed,
in particular in view of the time of flight of cosmic photons
in Gamma Ray Burst (GRBs). In a GRB photons over a range of about
$10^{5} \dots 10^{8}$ eV are emitted within a few seconds or minutes
from a small source. In particular, evaluating the times of photon
arrival for 35 GRBs against the dispersion ansatz
$E = | \vec p | + E/M$ (where $M$ is a large mass due to some
``quantum gravity foam''), Ref.\ \cite{EMNSS} concluded
$ M > 0.001 M_{\rm Planck}$. It has been proposed to establish
a bound on $\Vert \Theta \Vert$ by similar considerations \cite{Camel},
which should address the IR singularity in eq.\ (\ref{disp}).
However, the constant $C$ turned out to be {\em negative} to one 
loop \cite{LLT},
which suggested that NC QED is IR unstable and thus ill-defined.
In Ref.\ \cite{BNSV} we revisited
NC QED non-perturbatively: we discretised the commutative plane with
an $L\times L$ lattice and the NC plane with a TEK model of matrix size
$ N \approx L$. A physical scale was identified by matching Wilson
loop expectation values (and further observable) at different $\beta$
values. To a good approximation this led to $a \propto 1/ \beta$,
so that the DSL is taken at fixed $\, \theta \propto N / \beta^{2} \, $
(unlike NC QED$_{2}$), see Figure \ref{4dWl}.\footnote{The 
precise fine tuning
results, with slight deviation from this rule, are given in Ref.\
\cite{BNSV}.}

\begin{figure}
\vspace*{-3mm}
\begin{center}
\includegraphics[angle=270,width=.45\linewidth]{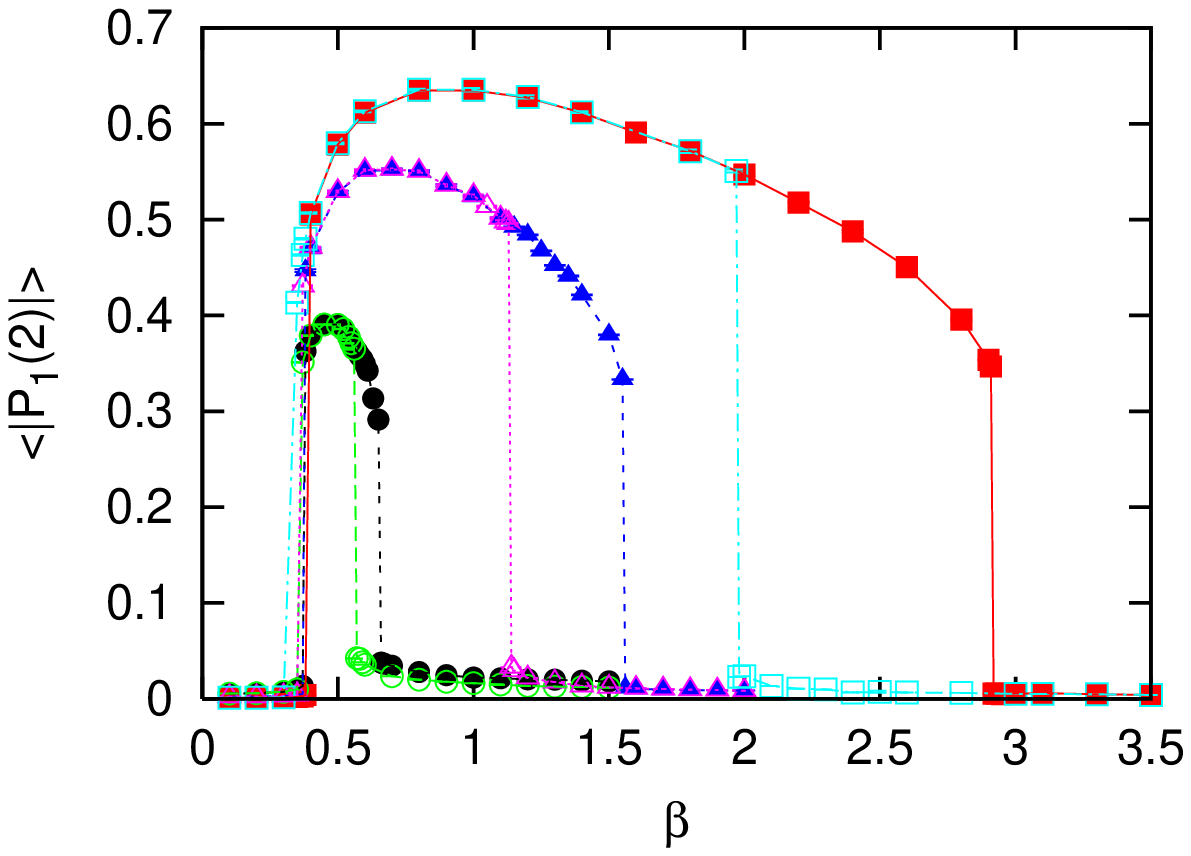}
\includegraphics[angle=270,width=.45\linewidth]{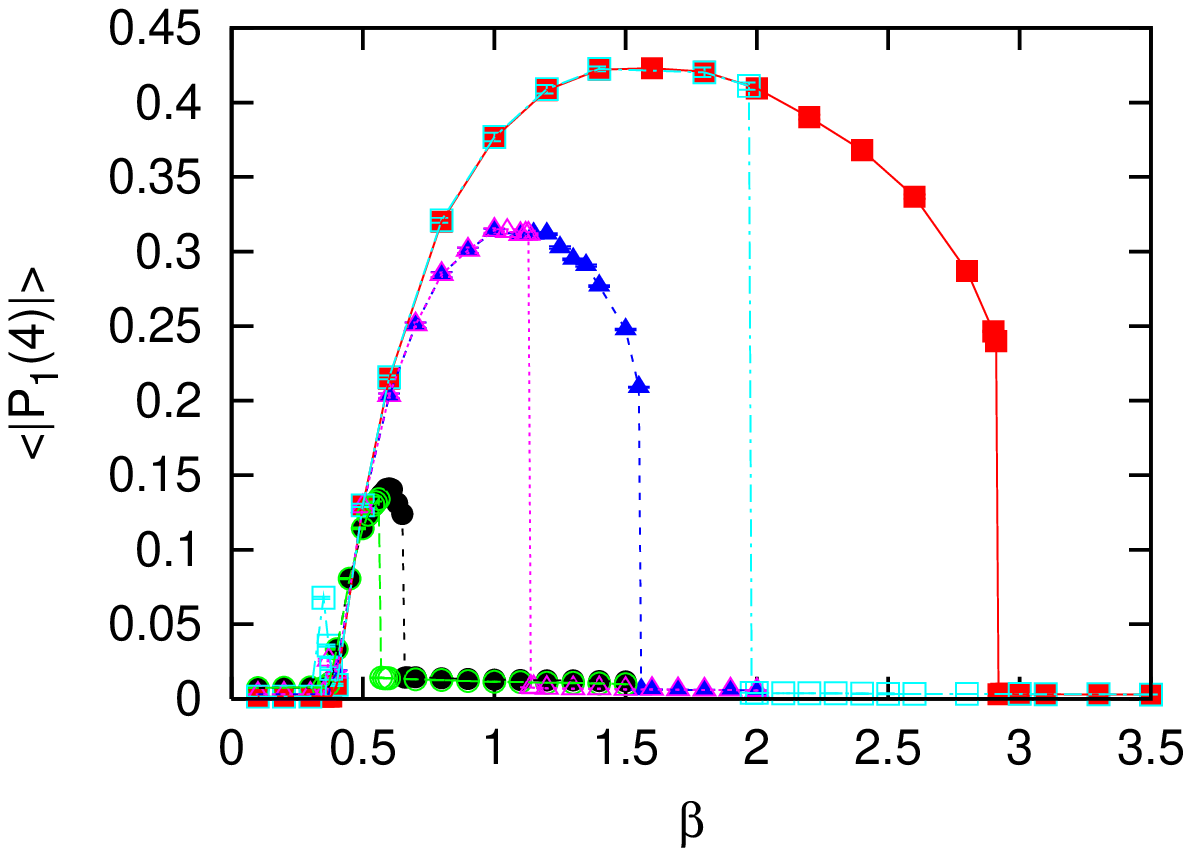}
\vspace*{-3mm} \\
\end{center}
\caption{{\it The expectation values of open Polyakov lines 
of length $2a$ (on the left) and $4a$ (on the right) for
$N=15, \ 25 $ and $35$ (curves from left to right, with hysteresis
at the transition to the weak coupling phase). At intermediate
gauge coupling we recognise a phase of broken translation symmetry.}}
\label{Polyfig}
\vspace*{-3mm}
\end{figure}

As an order parameter for translation symmetry in the NC plane we
measured the open Polyakov line, which is 
$\star$-gauge invariant and which carries momentum $p$,
\beq
P_{\mu}(n) = {\cal P} \exp_{\star}
\Big( ig \int_{x}^{x+ \tilde p_{\mu}} A_{\mu} (\xi ) \, d \xi_{\mu} \Big) \ ,
\qquad {\cal P}~:~{\rm path~ordering},~ 
\tilde p_{\mu} 
= na \hat \mu ~:~{\rm length} \ .
\eeq
Figure \ref{Polyfig} shows the (expected) symmetric phases at
strong and at weak coupling, but a broken phase above 
$\beta \simeq 0.35$. The upper end of the broken phase rises 
roughly $\propto N^{2}$, and its hysteresis behaviour implies that
the corresponding phase transition is of first order.
\footnote{The broken phase of the TEK model
was mentioned earlier \cite{Ishi} and later confirmed in Ref.\ \cite{Ox}.} 

\begin{figure}
\vspace*{-3mm}
\begin{center}
\includegraphics[angle=0,width=.62\linewidth]{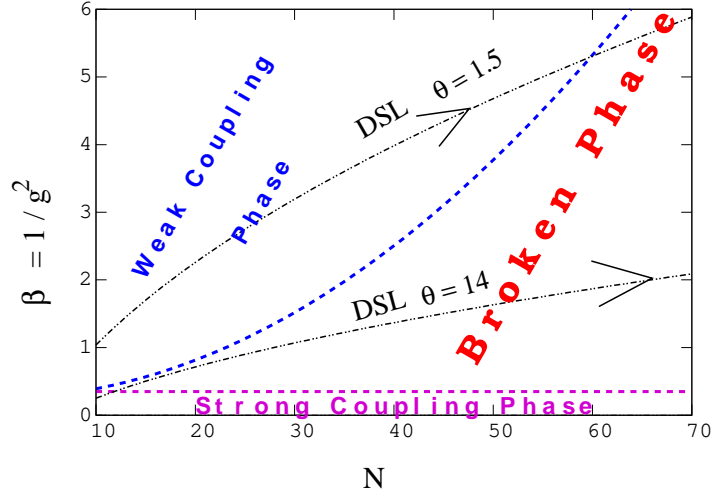}
\vspace*{-6mm} \\
\end{center}
\caption{{\it The phase diagram for QED$_{4}$ with a NC plane:
between the strong coupling phase ($\beta < 0.35$) and the weak
coupling phase we find a phase of broken translation invariance.
Since the weak/broken transition rises like $\beta \propto N^{2}$, whereas
the DSL for a fixed NC parameter $\theta$ follows $\beta \propto
\sqrt{N}$, the DSL always ends up in the broken phase.}} 
\vspace*{-1mm}
\label{phasedia}
\end{figure}
\begin{figure}
\vspace*{-3mm}
\begin{center}
\includegraphics[angle=270,width=.49\linewidth]{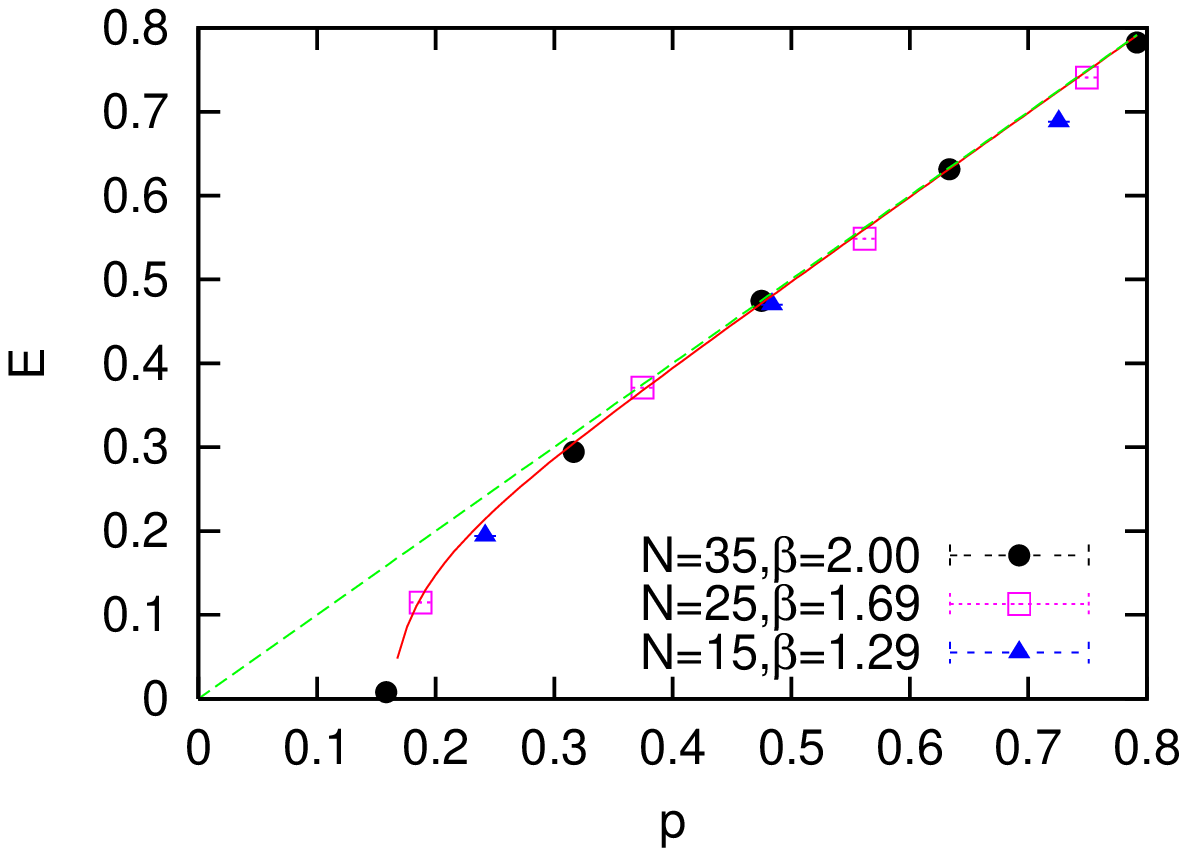}
\includegraphics[angle=270,width=.49\linewidth]{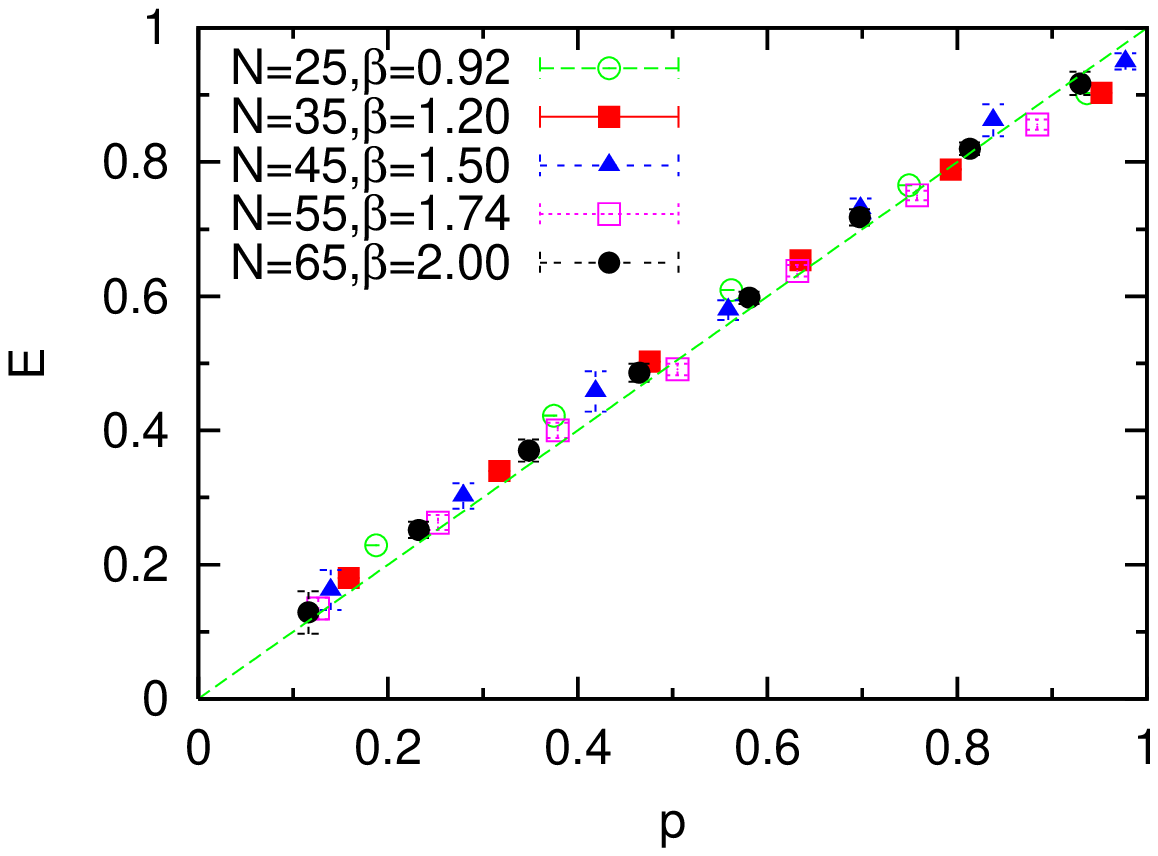}
\vspace*{-2mm} \\
\end{center}
\caption{{\it Dispersion relations $\, E(p) \ $
for the photon in a NC space in the weak coupling phase 
(on the left, $p=p_{1}$), 
and in the broken phase (on the right, $p=p_{3}$).
The former result is consistent with the perturbatively predicted IR
instability (``tachyonic'' behaviour, eq.\ (3.1) with $C\simeq 0.1285$). 
In the physical phase the photon 
is massless again, corresponding to a Nambu-Goldstone boson of the 
broken translation symmetry.}}
\label{disprel}
\vspace*{-3mm}
\end{figure}

This leads to the phase diagram \ref{phasedia}, where we also mark
the trajectories for DSLs with different values of $\theta$.
These DSL curves always lead to the broken phase, where we observe
IR stability of all observables measured \cite{BNSV}. The perturbative result
describes correctly the weak coupling phase, as the dispersion relation
(in the commutative plane) in Figure \ref{disprel} on the left shows. 
The plot on the right
refers to the broken phase, which captures the physically relevant
DSL. Here the photon dispersion is
linear, and the photon can be identified with the
Nambu-Goldstone boson of spontaneous translation symmetry breaking.
\footnote{This interpretation is known for instance from Ref.\ 
\cite{Kostel}. The ansatz taken there incorporates a NC space where
the star product is truncated in $O(\Vert \Theta \Vert)$
\cite{Kostel2}. In that case, however, locality is restored, so the theory
is altered qualitatively.}

\section{Conclusions}

We simulated QED$_{2}$ and QED$_{4}$ on spaces containing a NC plane.
In both cases we observed a stable behaviour in the DSL 
to a continuous NC space of infinite extent, 
which suggests 
renormalisability. UV/IR mixing is manifest as a non-perturbative effect,
in agreement with numerical DSL results for the 3d NC $\lambda \phi^{4}$
model \cite{phi4}.

$\bullet$ In $d=2$ the APD invariance of Wilson loops breaks, without
any residual subgroup. Hence there is hardly
hope for an analytic solution, but
we may hope for a rich structure to be explored numerically 
\cite{BHN,BBT}.

$\bullet$ In $d=4$ the DSL leads to a phase of intermediate coupling strength
and broken translation symmetry. This physical phase appears to be IR 
stable --- in contrast to the weak coupling phase. Therefore the NC space
may accommodate photons after all \cite{BNSV}.

\vspace*{1mm}

\noindent
{\bf Acknowledgement :} {\it This work was supported by the 
DFG, HPC-Europe, INFN and D.O.E.\
under cooperative research agreement
DE-FG02-05ER41360.
The computations were performed in part
on parallel machines at HLRN and HLRS,
and on a PC cluster at Humboldt-Universit\"{a}t.
}

\end{document}